\documentclass[conference,letterpaper]{IEEEtran}

\usepackage{silence}
\usepackage[backend=bibtex, style=numeric,sorting=none, maxnames=5]{biblatex}
\usepackage{macro}

\usepackage{balance}

\title{Neural Polar Decoders for Deletion Channels}
\author{\IEEEauthorblockN{Ziv Aharoni}
\IEEEauthorblockA{Duke University\\
Email: ziv.aharoni@duke.edu}}
\author{%
 \IEEEauthorblockN{Ziv Aharoni  and Henry D. Pfister}
\IEEEauthorblockA{Department of Electrical and Computer Engineering\\
                   Duke University\\
                   Email: \{ziv.aharoni,henry.pfister\}@duke.edu}}
\begin{document}

\maketitle
\glsdisablehyper
\begin{abstract}
This paper introduces a \gls{npd} for deletion channels with a constant deletion rate. Existing polar decoders for deletion channels exhibit high computational complexity of $O(N^4)$, where $N$ is the block length. This limits the application of polar codes for deletion channels to short-to-moderate block lengths. In this work, we demonstrate that employing \glspl{npd} for deletion channels can reduce the computational complexity. First, we extend the architecture of the \gls{npd} to support deletion channels. Specifically, the \gls{npd} architecture consists of four \glspl{nn}, each replicating fundamental \gls{sc} decoder operations. To support deletion channels, we change the architecture of only one. The computational complexity of the \gls{npd} is $O(AN\log N)$, where the parameter $A$ represents a computational budget determined by the user and is independent of the channel. We evaluate the new extended \gls{npd} for deletion channels with deletion rates $\delta\in\{0.01, 0.1\}$ and we verify the \gls{npd} with the ground truth given by the trellis decoder by Tal et al. We further show that due to the reduced complexity of the \gls{npd}, we are able to incorporate list decoding and further improve performance. We believe that the extended \gls{npd} presented here could have applications in future technologies like DNA storage.
\end{abstract}
\blfootnote{The work of Z. Aharoni and H. D. Pfister was supported in part by the National Science Foundation (NSF) under Grant Numbers 2212437 and 2308445. Any opinions, findings, recommendations, and conclusions expressed in this material are those of the authors and do not necessarily reflect the views of these sponsors. The code used to produce the results of this paper is available at: \url{https://github.com/zivaharoni/neural-polar-decoders-dna-storage}. This version is identical to the one presented at the IEEE International Symposium on Information Theory (ISIT) 2025, with only minor typographical corrections.}
\glsresetall
\section{Introduction}

\par The deletion channel is perhaps the simplest channel with synchronization errors and it is a common model for communication systems \cite{gallagerSequentialDecodingBinary1961, dobrushinShannonsTheoremsChannels1967} and DNA storage \cite{bancroftLongtermStorageInformation2001,churchNextgenerationDigitalInformation2012, goldmanPracticalHighcapacityLowmaintenance2013}. A binary-input deletion channel with a constant deletion rate $\delta\in(0,1)$ deletes each bit independently with probability $\delta$. As a result, a binary length-$N$ input sequence $\mathbf{X}$ is mapped to a binary length-$D$ output sequence $\mathbf{Y}$, where $D\sim\mathsf{Bin}(N,1-\delta)$. 
Despite its memoryless behavior, the loss of synchronization caused by deletions makes the deletion channel difficult to analyze. Still, communicating over deletion channels is a challenging task that has been researched extensively in recent years \cite{daveyReliableCommunicationChannels2001,varshamovCodesWhichCorrect1965a, zigangirovSequentialDecodingBinary1969,sloaneSingledeletioncorrectingCodes2002}.
Many known decoders for the deletion channel assume only a fixed small number of deletions have occurred. However, solutions for the deletion channel with a constant deletion rate are more scarce.

\par In recent years, polar codes \cite{arikanChannelPolarizationMethod2009} were applied to deletion channels in a series of papers \cite{thomasPolarCodingBinary2017,tianPolarCodesChannels2017,talPolarCodesDeletion2021}. The method in \cite{thomasPolarCodingBinary2017} used the \gls{sc} decoder for memoryless channels, as introduced in \cite{arikanChannelPolarizationMethod2009}, while considering all possible deletion patterns of the received sequence $\mathbf{Y}$. This is highly inefficient as the number of deletion patterns grows exponentially.  
The paper \cite{tianPolarCodesChannels2017} devises a \gls{sc} polar decoder by considering deletion patterns throughout the decoding process. The deletion patterns are coded into a trellis which is embedded in the \gls{sc} decoding algorithm to compute the posteriors of the synthetic channels. The computational complexity of the decoder is $O(d^3N\log N)$ for $d$ deletions. This is efficient for small value of $d$, as the authors demonstrated the decoders performance for low deletion rate of $\delta=0.002$. However, this decoder still suffers from high computational complexity when the deletion rate is larger. 

\par The method proposed in \cite{talPolarCodesDeletion2021} represents the joint distribution of $\mathbf{X},\mathbf{Y}$ using a trellis. 
For a given output sequence $ \mathbf{y} $, a trellis is constructed with $ N $ sections, each corresponding to an input symbol $ X_i $. Each section contains $ D $ nodes, and each node can have up to three outgoing edges connecting it to nodes in the adjacent section.
Using this trellis, a \gls{sc} decoder is defined that allows the exact computation of the posteriors for the synthetic channels. This enables one to estimate the \gls{mi} between $\mathbf{X},\mathbf{Y}$, and subsequently, design a \gls{sc} decoder. The trellis decoder for deletion channels is a useful tool but the decoding complexity is still $O(N^4)$. This weakness motivates us to devise an \gls{sc} decoder with near-optimal performance and reduced complexity.

\par Our main observation about the trellis in \cite{talPolarCodesDeletion2021} is that it is sparse, namely, most of the edges in the trellis are not used. This led us to find a way to ``compress" the trellis representation and create an \gls{sc} decoder that operates on the compressed trellis representation. To do this, we build upon \cite{aharoniDatadrivenNeuralPolar2024}, which introduces a \gls{npd} that shares the same structure as the \gls{sc} decoder but replaces its elementary operations by \glspl{nn}. In that work, embedding \glspl{nn} in the structure of the \gls{sc} decoder was shown to perform well on channels with memory, in the sense that it enabled the computation of the posterior of the synthetic channels consistently\footnote{Given sufficient samples of sequences $\mathbf{X},\mathbf{Y}$, the \gls{npd} can approximate the posterior with bounded estimation error.}. 

\par There are two main benefits of \glspl{npd}. First, a \gls{npd} does not require an explicit channel model; it only requires the channel as a black-box for generating samples of input-output pairs. Second, the computational complexity is determined by the parameterization of the \glspl{nn} that construct the \gls{npd}, and not by the state space size of the channel. For example, consider \glspl{fsc} which are a common model for channels with memory. \Glspl{fsc} have a corresponding \gls{sc} decoder, called \gls{sct} decoder \cite{wangConstructionPolarCodes2015}. The \gls{sct} decoder for \glspl{fsc} has computational complexity of $O(M^3N\log N)$, where $M$ is the number of channel states. The \gls{npd} with a computation budget $A$, however, has a computational complexity of $O(A N\log N)$, regardless of the number of channel states. 
In the context of deletion channels, the \gls{sct} decoder for \glspl{fsc} is important since despite the differences between \glspl{fsc} and deletion channels, their corresponding \gls{sc} decoders for polar codes share a common structure and attributes. 

\par Henceforth, this work focuses on adapting \glspl{npd} to deletion channels. The adaptation of \glspl{npd} to deletion channels involves updating only one \gls{nn} out of the four \glspl{nn} that compose the \gls{npd}. To evaluate the \gls{npd}, we compare its decoding errors with the decoding errors of the trellis decoder for deletion channels. Our results demonstrate that while the \gls{npd} is comparable with the trellis decoder in terms of decoding errors, it exhibits computational complexity that is lower by two orders of magnitude. We also show that reducing the complexity enables us to incorporate list decoding \cite{talListDecodingPolar2015}. This is an important advantage since list decoding in polar codes is crucial to approach maximum-likelihood decoding in the short-to-moderate block lengths, but computationally intractable for the trellis decoder for deletion channels.

\section{Background}
This section starts by introducing notation and basic definitions, particularly of a binary \gls{iid} deletion channel, which is henceforth called the deletion channel. It proceeds by presenting polar codes and \glspl{npd}.

\par We denote by $(\Omega,\cF,\bP)$ the underlying probability space on which all \glspl{rv} are defined. Sets are denoted by calligraphic letters, e.g. $\mathcal{X}$. 
We use the notation $\mathbf{X}$ to denote the random vector $(X_1,X_2,\dots,X_N)$ and $\mathbf{x}$ to denote its realization, where $N$ will be given by the context. The term $P_{\mathbf{X}}$ denotes the distribution of $\mathbf{X}$.
Throughout the paper, we assume that $\cX=\{0,1\}$.
We denote by $[N]$ the set of integers $\{1,\dots,N\}$. Let $\cD^c$ be the complement of the set $\cD$. 
Let $\mathbf{x}$ be a length-$N$ binary sequence. The support of $\mathbf{x}$ is defined by $\operatorname{supp}(\mathbf{x}) = \{ i \mid x_i = 1, \;  i \in [N] \}$.
Given an index set $\mathcal{D} \subseteq [N]$, the notation $\mathbf{X}_{\mathcal{D}}$ represents the subsequence of $\mathbf{X}$ consisting of the elements indexed by $\mathcal{D}$, i.e., $\mathbf{X}_{\mathcal{D}} = (X_i \mid i \in \mathcal{D})$ with the elements of $\cD$ in increasing order. 
The term $A\otimes B$ denotes the Kronecker product of $A$ and $B$ when $A,B$ are matrices, and it denotes the product distribution whenever $A,B$ are distributions. 
The term $A^{\otimes N}:=A\otimes A\otimes\dots\otimes A$ denotes an application of the $\otimes$ operator $N$ times. 
The notation $\mathbf{D}\sim \mathsf{Ber}(p)^{\otimes N}$ indicates that $\mathbf{D}$ is a length-$N$ \gls{iid} Bernoulli random vector with success rate $p\in(0,1)$.
The binary entropy of $X$ is denoted by $\f{\sH_2}{X}$ and the \gls{ce} between two distributions $P,Q$ is denoted by $\ce{P}{Q}$. The deletion channel is defined next.

\begin{definition}[Deletion Channel]\label{def:del}
Let $\delta\in(0,1)$ denote the deletion probability and let $N$ denote the block length. Let $\mathbf{D}\sim\mathsf{Ber}(\delta)^{\otimes N}$, $\cD=\operatorname{supp}(\mathbf{D})$ and $D=|\cD^c|$. Then, the output of the deletion channel for an input $\mathbf{X}$ is given by $\mathbf{Y} = \mathbf{X}_{\cD^c}$.
\end{definition}

\subsection{Polar Codes}
\par Let $G_N = B_N F^{\otimes n}$ be Ar\i{}kan's polar transform with the generator matrix for block length $N=2^n$ for $n\in\bN$. The matrix $F$ is Ar\i{}kan's $2\times 2$ kernel defined by $F=\begin{bmatrix}
    1 & 0 \\
    1 & 1
\end{bmatrix}.$
The matrix $B_N$ is the permutation matrix associated with the bit-reversal permutation. It is defined by the recursive relation $B_N = R_N(I_2 \otimes B_{\frac{N}{2}})$ starting from $B_2=I_2$.
The term $I_N$ denotes the identity matrix of size $N$ and $R_N$ denotes a permutation matrix called reverse-shuffle \cite{arikanChannelPolarizationMethod2009}. 
Given $\mathbf{x}\in \bF_2^N$ the polar transform is defined by 
\begin{equation}
    \mathbf{u}=\mathbf{x} G_N.
\end{equation}
The synthetic channels of the polar code are defined by 
\begin{equation}\label{eqn:synthetic_channels}
W^{}_{i,N}(u_i|u^{i-1},y^N) =  P_{U_i|U^{i-1},Y^N}(u_i|u^{i-1},y^N),
\end{equation}
for $i\in[N]$, $\mathbf{u}\in\cX^N$ and $\mathbf{y}\in\cY^N$. In this work, we set $\cX=\cY$, though in general, $\cY$ can be arbitrary.

\par A \gls{sc} polar decoder is composed of 4 elementary functions that map the channel outputs $\mathbf{y}$ into the estimation $\hat{\mathbf{u}}$ via the computation of the synthetic channels \cite{arikanChannelPolarizationMethod2009,aharoniDatadrivenNeuralPolar2024}. Thus, a \gls{sc} polar decoder may be defined as follows.
\begin{definition}[Successive cancellation polar decoder]\label{def:sc_polar_decoder}
Let $\mathbf{x}\in\cX^N$ and $\mathbf{y}\in\cY^N$. A \gls{sc} polar decoder is composed by the following functions:
\begin{itemize}
    \item The embedding function $E:\cY\to\cE$,
    \item The check-node function $F:\cE\times\cE\to\cE$,
    \item The bit-node function $G:\cE\times\cE\times\cX\to\cE$,
    \item The embedding-to-llr function $H:\cE\to\bR$,
\end{itemize}
where $\cE\subset\bR^d,\;d\in\bN,$ is the embedding space of the \gls{sc} decoder.
\end{definition}

\par Definition \ref{def:sc_polar_decoder} lists the ingredients needed by a \gls{sc} decoder to decode an output sequence $\mathbf{y}\in\cX^{N\times 1}$ into $\hat{\mathbf{u}}\in\cX^{N\times 1}$. The decoding computation starts with mapping each $y\in\mathbf{y}$ independently into $e = E(y)$, $e\in\bR^{1\times d}$, to form $\mathbf{e}\in\bR^{N\times d}$. The decoding computation is completed by using $\mathbf{e}$ and $F,G,H$ to compute $\hat{\mathbf{u}}$ via the recursion of the \gls{sc} decoder, as shown in \cite{arikanChannelPolarizationMethod2009}. For example, setting $\cE=\bR$ and defining $E,F,G,H$ by 
\begin{align}\label{eqn:scd_memoryless}
    E(y) &= \log\frac{\f{W}{y|1}}{\f{W}{y|0}}+ \log\frac{\f{P_{X}}{1}}{\f{P_{X}}{0}}, \nn\\
    F(e_1, e_2) &=  -2\tanh^{-1}\left(\tanh{\frac{e_1}{2}}\tanh{\frac{e_2}{2}}\right), \nn\\
    G(e_1, e_2, u) &=  e_2 + (-1)^{u}e_1, \nn\\
    H(e_1) &= e_1,
\end{align}
for $e_1,e_2\in\bR, u\in\cX$, defines a \gls{sc} decoder for memoryless channels. For \glspl{fsc} with state space $\cS$, the \gls{sct} decoder from \cite{wangConstructionPolarCodes2015} is defined by setting $\cE=[0,1]^{|\cX||\cS|^2}$ and defining 
\begin{align}\label{eqn:scd_fscs}
    \left[E(y)\right]_{x,s_0,s_1} = \f{P_{X,Y,S^\prime|S}}{x,y,s_1|s_0}.
\end{align}
The coordinates $x,s_0,s_1$ are used to specify an entry of the vector $e$, and $y$ is the input of the function $E$.
Here, the embedding function outputs a vector, mostly referred as a trellis, that captures the joint distribution of $X,Y$ and the channel states before and after the transmission $S,S^\prime$, respectively. The functions $F,G,H$ are given in \cite{wangConstructionPolarCodes2015} and omitted here due to space limitations.

\subsection{Neural Polar Decoders}
\par An \gls{npd} is also an instance of a \gls{sc} polar decoder, as in Definition \ref{def:sc_polar_decoder}. Setting the $\cE=\bR^d,\;d\in\bN$ and realizing the functions $E,F,G,H$ by \glspl{nn} yields an \gls{npd}. Let $\cG_\mathsf{NN}^{(d_i,k,d_o)}$ denote the family of shallow \glspl{nn} with $d_i$ inputs, $k$ hidden units, $d_o$ outputs and ReLU activations. An \gls{npd} is defined as follows.
\begin{definition}[Neural polar decoder]\label{def:npd}
    An \gls{npd} is an \gls{sc} polar decoder, as defined in Definition \ref{def:sc_polar_decoder}, with the functions $E,F,G,H$ defined by:
    \begin{itemize}
        \item The embedding \gls{nn} $E_{\theta_E}\in\cG_\mathsf{NN}^{(1,h,d)}$,
        \item The check-node \gls{nn} $F_{\theta_G}\in\cG_\mathsf{NN}^{(2d,h,d)}$,
        \item The bit-node \gls{nn} $G_{\theta_G}\in\cG_\mathsf{NN}^{(2d+1,h,d)}$,
        \item The embedding-to-llr \gls{nn} $H_{\theta_H}\in\cG_\mathsf{NN}^{(d,h,1)}$,
    \end{itemize}
    where $h\in \bN$ is the number of hidden units. Let $\theta= \left\{\theta_E,\theta_F,\theta_G,\theta_H\right\}$ be the parameters of the \gls{npd} and $\Theta$ the parameter space.
\end{definition}
For any $\theta\in\Theta$ the \gls{npd} computes an estimate of the posterior distribution of the synthetic channels $W_{i,N}$, as given in \eqref{eqn:synthetic_channels} via the recursion of the \gls{sc} decoder. Let this estimate be denoted by $W^\theta_{i,N}$. The process of determining the \gls{npd}'s parameters is a \gls{sgd} optimization process in which examples of input output pairs $\mathbf{x},\mathbf{y}\stackrel{iid}{\sim}P_{\mathbf{X},\mathbf{Y}}$ are used to optimize the parameters of the \gls{npd}, $\theta$. The optimization process minimizes the \gls{ce} $\ce{W_{i,N}}{W^\theta_{i,N}}$, which achieves its minimum of $\f{\sH_2}{W_{i,N}}$ if and only if $W_{i,N}=W^\theta_{i,N}$ almost surely. The optimization algorithm is outlined in \cite{aharoniDatadrivenNeuralPolar2024}.

\section{Neural Polar Decoder for Deletion Channels}
This section addresses the adaptation of \glspl{npd} to deletion channels. It starts by discussing the limitations of current \glspl{npd} regarding deletion channels. It proceeds by adapting the \gls{npd} to deletion channels with a focus on the embedding \gls{nn} $E_\theta$.  

\subsection{Limitation of current NPDs on deletion channels}
The \gls{npd} in \cite{aharoniDatadrivenNeuralPolar2024} is designed for channels with synchronized inputs and outputs, i.e. $\mathbf{x},\mathbf{y}$ are always the same length $N$ and $y_i$ is an observation of $x_i$. However, in the deletion channel, this is not the case. Therefore, the \gls{npd} must be changed in order to accommodate deletion channels.

\subsection{Adapting the embedding NN}
To solve the problem that $\mathbf{x}\in\cX^{N\times 1}$ and $\mathbf{y}\in\cX^{D\times 1}$ do not have the same length, we adapt the embedding \gls{nn} from Definition \ref{def:npd}. Instead of computing $\mathbf{e}\in\bR^{N\times d}$ by using the embedding \gls{nn} to compute $e = E_\theta(y)$ for each $y\in\mathbf{y}$, we instead compute $\mathbf{e}=\mathbf{E}_\theta(\mathbf{y})$, where $\mathbf{E}_\theta$ is a \gls{nn} that maps the whole $\mathbf{y}$ vector to $\mathbf{e}$. It is possible to build an \gls{npd} for every possible output length $D$. In the next section, we show that this approach yields a consistent estimation of the synthetic channels, and therefore defines a valid \gls{sc} decoder. However, this approach is not efficient in parameters.
\subsection{Consistency of Neural polar decoders on deletion channels}
The following theorem shows that it is possible to build a different \gls{npd} for every output length $D$.
\begin{theorem}\label{thm:nsc}
    Let $\mathbf{X}\in\cX^N$, $N=2^n$, $n\in \bN$, be an input of a deletion channel, as given in Definition \ref{def:del}. Let $\mathbf{Y}\in\cX^D$ be the output, where $D\le N$ is a constant parameter. Let $\cS_M=\left\{\mathbf{x}^{(j)},\mathbf{y}^{(j)}\right\}$ be $M\in\bN$ independent observations of $\mathbf{X},\mathbf{Y}$. Let  $\mathbf{u}^{(j)} = \mathbf{x}^{(j)}G_N$. Then, for every $\varepsilon>0$ there exists $p\in\bN$, compact $\Theta \in\bR^p$ and $m\in\bN$ such that for $M>m$ and $i\in[N]$, $\bP-a.s.$
    \begin{align}
        \left| \f{\sH_{\Theta}^M}{U_i\vert U_1^{i-1}, \mathbf{Y}} -\f{\sH}{U_i\vert U_1^{i-1}, \mathbf{Y}}\right| < \varepsilon,
    \end{align}
    where,
    \begin{align}
        \f{\sH_{\Theta}^M}{U_i\vert U_1^{i-1}, \mathbf{Y}} = \min_{\theta\in\Theta} -\frac{1}{M}\hspace{-0.5em}\sum_{\mathbf{x},\mathbf{y}
        \in\cS_M} \log \f{W^\theta_{i,N}}{u_i\vert u^{i-1},\mathbf{y}}
    \end{align}
\end{theorem}
\noindent The proof of Theorem \ref{thm:nsc} is similar of the proof of \cite[Theorem 3]{aharoniDatadrivenNeuralPolar2024} and is omitted here due to space limitations. 

\subsection{One Embedding NN for all $D$}


\par The main goal now is to describe the architecture of the embedding \gls{nn}, such that it will be applicable for all $D$. The embedding \gls{nn} for the deletion channel is defined next. 
\begin{definition}[Embedding NN for the deletion channel]
Let $\mathbf{y}\in \cX^D$ be the channel outputs. The embedding \gls{nn} is defined by the following equations:
\begin{align*}
    \mathbf{\tilde{y}} &= (\mathbf{y},\mathbf{?}^{N-D}), && \text{(pad with erasure symbols)}\\
    \mathbf{\tilde{e}} &= \operatorname{Embedding}(\mathbf{\tilde{y}}), && \text{(embed channel outputs)}\\
    \mathbf{\tilde{e}}_\mathsf{pos} &= \mathbf{\tilde{e}} + \mathbf{p}, && \text{(positional encoding)}\\
    \mathbf{e} &= \mathsf{CNN}(\mathbf{\tilde{e}}_\mathsf{pos}), && \text{(Conv. NN)}
\end{align*}
where $(\mathbf{y},\mathbf{?}^{N-D})$ is a concatenation of an erasure symbol $?$ to pad $\mathbf{y}$ to match with the input block length $N$. 
The embedding layer, denoted as $\operatorname{Embedding}$, maps each value in $\widetilde{\mathcal{Y}} = \{0,1,?\}$ to a unique vector in $\mathbb{R}^{d}$. The embedded values $\mathbf{\tilde{e}}\in\bR^{N\times d}$ are then encoded with positional encoding matrix $\mathbf{p}\in \bR^{N\times d}$, as given in \cite{vaswaniAttentionAllYou2017}. For $i\in[N], j\in[d]$
\begin{align*}
    p_{i, 2j} &= \sin\left(\frac{i}{10000^{\frac{2j}{d}}}\right), \\
    p_{i, 2j+1} &= \cos\left(\frac{i}{10000^{\frac{2j}{d}}}\right).
\end{align*}
Lastly, $ \mathbf{\tilde{e}}_\mathsf{pos}$ is processed in convolutional \gls{nn} to yield the embedding vectors used by the \gls{npd}.
\end{definition}




\section{Experiments}
This section presents the experiments conducted on deletion channels. In the experiments, we tested performance of the \gls{npd} and compared it to the ground truth obtained by the trellis decoder in \cite{talPolarCodesDeletion2021}. 

\subsection{Setup}
\begin{figure}[b!]
    \centering
    \begin{tikzpicture}
    \begin{axis}[
        width=8cm, height=8cm,
        xlabel=\textcolor{dimgray85}{$\log_2 N$},
        ylabel=\textcolor{dimgray85}{Decoder speed (blocks/sec)},
        symbolic x coords={5, 6, 7, 8, 9},
        xtick=data,
        ytick={0},
        ybar=2pt,
        bar width=10pt,
        ymax=3750,
        legend pos=north east,
        legend style={font=\small},
        enlarge x limits=0.12,
        y tick label style={rotate=90},
        every node near coord/.append style={font=\small, rotate=90, anchor=west}, 
        nodes near coords={\pgfmathprintnumber[fixed, precision=3]{\pgfplotspointmeta}}, 
    nodes near coords align={vertical},
    ]
    \addplot+ [bar shift=0pt] coordinates {(5, 30.16) (6, 10.74) (7, 2.67) (8, 0.92) (9, 0.16)};
    \addplot+ [bar shift=-12pt] coordinates {(5, 8.39) (6, 1.83) (7, 0.26) (8, 0.02) (9, 0.001)};
    \addplot+ [bar shift=12pt] coordinates {(5, 2907) (6, 1470) (7, 757) (8, 367) (9, 185)};
    \legend{SCT $\delta=0.01$, SCT $\delta=0.1$, NPD}
    \end{axis}
    \end{tikzpicture}
    \caption{Comparison of decoder speed of the \gls{npd} and the \gls{sct} for deletion channels.}
    \label{fig:complexity}
\end{figure}
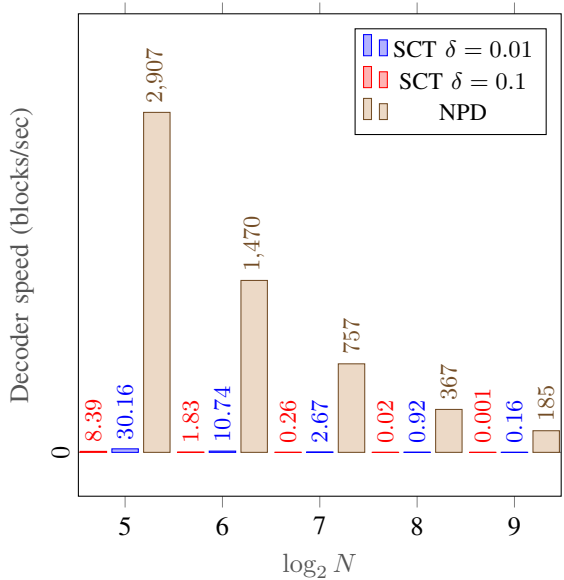

\par The experiments were conducted using block lengths ranging from $N=32$ to $N=512$. The input distribution was chosen to be uniform and \gls{iid}, with $P_{\mathbf{X}} = \mathsf{Ber}(\frac{1}{2})^{\otimes N}$. The deletion channel rates were set to $\delta\in\{0.01, 0.1\}$. We compared the trellis decoder from \cite{talPolarCodesDeletion2021} with our proposed method. The performance metrics evaluated include \gls{fer} and running time. 

\par Additionally, we evaluated the \gls{npd} with \gls{scl} decoding \cite{talListDecodingPolar2015}. However, due to the high computational complexity of the trellis decoder, we were unable to evaluate it with \gls{scl}, particularly for moderate block lengths, even without \gls{scl} decoding.

\par In all experiments, the \gls{npd} was trained on $5,000,000$ independent input-output pairs of $\mathbf{X},\mathbf{Y}$. For each block length $N\in\{32,64,128,256,512\}$, a separate \gls{npd} was trained. After the training procedure was completed, the parameters of the \gls{npd} were fixed and used for decoding.

\subsection{Numerical Results}
This section presents the numerical results conducted to evaluate \glspl{npd} on deletion channels. The results focus on the decoding complexity and the decoding performance measured by \gls{fer}.
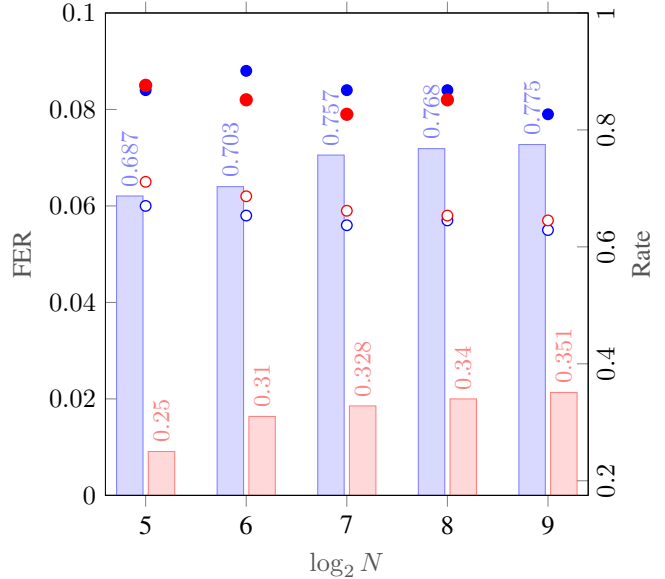
\begin{figure}[t!]
    \centering
    \begin{tikzpicture}
    \begin{axis}[
        width=8cm, height=8cm,
        xlabel=\textcolor{dimgray85}{$\log_2 N$},
        ylabel=\textcolor{dimgray85}{Rate},
        symbolic x coords={5, 6, 7, 8, 9},
        xtick=data,
        ybar=2pt,
        bar width=10pt,
        ymax=1.0,
        legend pos=north east,
        legend style={font=\small},
        axis y line*=right,
        y tick label style={rotate=90},
        every node near coord/.append style={font=\small, rotate=90, anchor=west}, 
        nodes near coords={\pgfmathprintnumber[fixed, precision=3]{\pgfplotspointmeta}}, 
    nodes near coords align={vertical},
    ]
    \addplot+[opacity=0.5] coordinates {(5, 0.687) (6, 0.703) (7, 0.757) (8, 0.768) (9, 0.775)};
    \addplot+[opacity=0.5] coordinates {(5, 0.25) (6, 0.31) (7, 0.328) (8, 0.34) (9, 0.351)};
    \end{axis}
\begin{axis}[
    width=8cm, height=8cm,
    ylabel=\textcolor{dimgray85}{FER},
    xlabel={}, 
    xtick=\empty,
    ytick pos=left,
    ymin=0,
    ymax=0.1,
    yticklabel style={/pgf/number format/fixed,/pgf/number format/precision=2}, 
    legend pos=north east,
    legend style={font=\small}
]
\addplot[only marks, mark=*, color=blue, mark options={fill=blue, draw=blue, line width=0.5pt, scale=1}] coordinates {(5, 0.084) (6, 0.088) (7, 0.084) (8, 0.084) (9, 0.079)};
    \addplot[only marks, mark=*, mark options={fill=white, draw=blue, line width=0.5pt, scale=1}] coordinates {(5, 0.06) (6, 0.058) (7, 0.056) (8, 0.057) (9, 0.055)};
\addplot[only marks, mark=*, color=red, line width=1pt, fill=none] coordinates {(5, 0.085) (6, 0.082) (7, 0.079) (8, 0.082) };
    \addplot[only marks, mark=*, mark options={fill=white, draw=red, line width=0.5pt, scale=1}] coordinates {(5, 0.065) (6, 0.062) (7, 0.059) (8, 0.058) (9, 0.057)};
\end{axis}
\end{tikzpicture}
    \caption{Comparison of the SCT and the NPD for various block lengths. The bar plot shows the information rate per block length. The marks shows the attained \glspl{fer}. Blue/red plots correspond to $\delta=0.01,0.1$, respectively.}
    \label{fig:fer-sc}
\end{figure}
\subsubsection{Computational complexity}
\par Figure \ref{fig:complexity} compares the decoder's speed, measured in blocks-per-second, of the \gls{npd} and the \gls{sct} decoder from \cite{talPolarCodesDeletion2021}. Results are shown for both $\delta=0.1$ and $\delta=0.01$ because the complexity of \gls{sct} decoding depends on the number of deletions. The computational complexity of the \gls{npd}, however, depends on the parametrization of the \gls{npd} and not on the channel. Therefore, its computational complexity is the same for both values of $\delta$. It is clear from the figure that the decoding speed of the \gls{npd} is significantly faster than the \gls{sct} decoder.

\begin{figure}[t!]
    \centering
    \begin{tikzpicture}
        \begin{axis}[
        width=8cm, height=8cm,
        ylabel=\textcolor{dimgray85}{Rate},
        symbolic x coords={5, 6, 7, 8, 9},
        xtick=data,
        ybar=2pt,
        bar width=10pt,
        ymax=0.8,
        axis y line*=right,
        xlabel={}, 
        xtick=\empty,
        ytick=\empty,
        ytick pos=right,
        y tick label style={rotate=90},
        every node near coord/.append style={font=\small, rotate=90, anchor=west}, 
        nodes near coords={\pgfmathprintnumber[fixed, precision=3]{\pgfplotspointmeta}}, 
    nodes near coords align={vertical},
    ]
    \addplot+[opacity=0.35] coordinates {(5, 0.687) (6, 0.703) (7, 0.757) (8, 0.768) (9, 0.775)};
    \end{axis}
    \begin{axis}[
        width=8cm, height=8cm,
        xlabel=\textcolor{dimgray85}{$\log_2 N$},
        ylabel=\textcolor{dimgray85}{FER},
        symbolic x coords={5, 6, 7, 8, 9},
        xtick=data,
        ymax=0.1,
        legend style={at={(0.5,1.0)},
                      font=\small,
                      anchor=south,},
        legend columns=2,
        ytick pos=left,
        yticklabel style={/pgf/number format/fixed,/pgf/number format/precision=2}, 
    nodes near coords align={vertical},
    ]
    \addplot[mark=*, color=black, mark options={fill=black, draw=black, line width=0.5pt, scale=1}] coordinates {(5, 0.084) (6, 0.088) (7, 0.084) (8, 0.084) (9, 0.079)};
    \addplot[mark=*, color=darkblue, mark options={fill=white, draw=darkblue, line width=0.1pt, scale=1}] coordinates {(5, 0.06) (6, 0.058) (7, 0.056) (8, 0.057) (9, 0.055)};
    \addplot[mark=square*, color=darkred, mark options={fill=white, draw=darkred, line width=0.1pt, scale=1}] coordinates {(5, 0.026) (6, 0.025) (7, 0.027) (8, 0.029) (9, 0.0296)};
    \addplot[mark=triangle*, color=darkgreen, mark options={fill=white, draw=darkgreen, line width=0.1pt, scale=1}] coordinates {(5, 0.023) (6, 0.024) (7, 0.0206) (8, 0.024) (9, 0.0233)};
    \addplot[mark=triangle*, color=darkorange, mark options={rotate=180,fill=white, draw=darkorange, line width=0.1pt, scale=1}] coordinates {(5, 0.022) (6, 0.0237) (7, 0.021) (8, 0.023) (9, 0.0229)};
    \legend{SCT $L=1$, NPD $L=1$, NPD $L=2$,NPD $L=4$,NPD $L=8$,}
    \end{axis}
\end{tikzpicture}
    \caption{\Glspl{fer} attained by SCL decoding of the \gls{npd} for  $\delta=0.01$. The bar plot shows the information rate per block length.}
    \label{fig:fer-scl-01}
\end{figure}
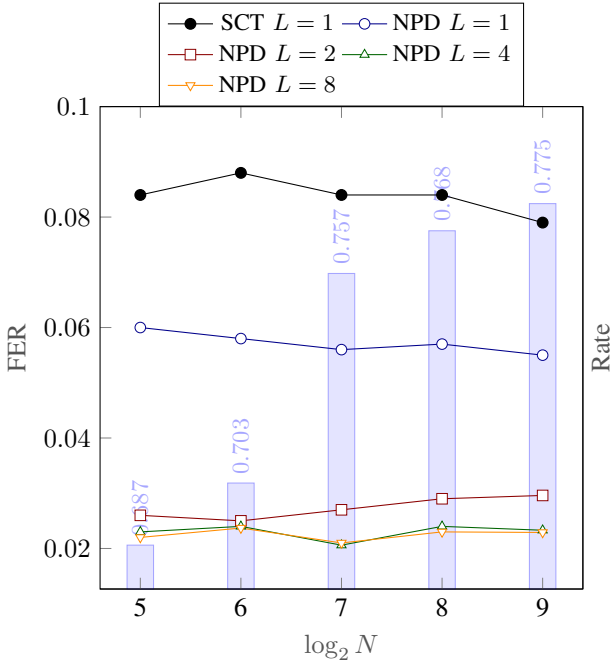
\subsubsection{Frame error rate of SC decoding}
\par Figure \ref{fig:fer-sc} compares the \gls{sct} decoder and the \gls{npd} by illustrating the \glspl{fer} attained by both decoders for various block lengths. The blue and red dots and bars correspond to $\delta=0.01$ and $\delta=0.1$, respectively. The codes were designed to achieve \gls{fer} $\sim 0.1$; the bars indicate the information rate used per block length. The solid marks show the \glspl{fer} attained by the \gls{sct} decoder and the hallow marks show the \glspl{fer} attained by the \gls{npd}. The results of the \gls{npd} are slightly better, even though, theoretically, the results of the \gls{npd} should not improve. We conjecture that this is due to numerical errors in the implementation of the \gls{sct} algorithm in \cite{talPolarCodesDeletion2021}.
\subsubsection{Frame error rate of SCL decoding}
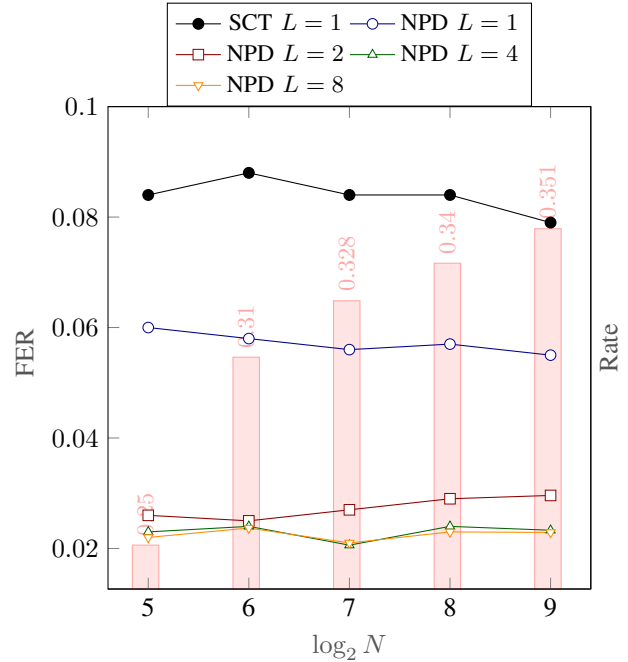
\begin{figure}[t!]
    \centering
    \begin{tikzpicture}
        \begin{axis}[
        width=8cm, height=8cm,
        ylabel=\textcolor{dimgray85}{Rate},
        symbolic x coords={5, 6, 7, 8, 9},
        xtick=data,
        ybar=2pt,
        bar width=10pt,
        ymax=0.39,
        axis y line*=right,
        xlabel={}, 
        xtick=\empty,
        ytick=\empty,
        ytick pos=right,
        y tick label style={rotate=90},
        every node near coord/.append style={font=\small, rotate=90, anchor=west}, 
        nodes near coords={\pgfmathprintnumber[fixed, precision=3]{\pgfplotspointmeta}}, 
    nodes near coords align={vertical},
    ]
    \addplot+[opacity=0.0] coordinates {(5, 0.687) (6, 0.703) (7, 0.757) (8, 0.768) (9, 0.775)};
    \addplot+[bar shift=-1pt, opacity=0.35] coordinates {(5, 0.25) (6, 0.31) (7, 0.328) (8, 0.34) (9, 0.351)};
    \end{axis}
    \begin{axis}[
        width=8cm, height=8cm,
        xlabel=\textcolor{dimgray85}{$\log_2 N$},
        ylabel=\textcolor{dimgray85}{ FER },
        symbolic x coords={5, 6, 7, 8, 9},
        xtick=data,
        ymax=0.1,
        legend style={at={(0.5,1.0)},
                      font=\small,
                      anchor=south,},
        legend columns=2,
        ytick pos=left,
        yticklabel style={/pgf/number format/fixed,/pgf/number format/precision=2}, 
    nodes near coords align={vertical},
    ]
    \addplot[mark=*, color=black, mark options={fill=black, draw=black, line width=0.5pt, scale=1}] coordinates {(5, 0.084) (6, 0.088) (7, 0.084) (8, 0.084) (9, 0.079)};
    \addplot[mark=*, color=darkblue, mark options={fill=white, draw=darkblue, line width=0.1pt, scale=1}] coordinates {(5, 0.06) (6, 0.058) (7, 0.056) (8, 0.057) (9, 0.055)};
    \addplot[mark=square*, color=darkred, mark options={fill=white, draw=darkred, line width=0.1pt, scale=1}] coordinates {(5, 0.026) (6, 0.025) (7, 0.027) (8, 0.029) (9, 0.0296)};
    \addplot[mark=triangle*, color=darkgreen, mark options={fill=white, draw=darkgreen, line width=0.1pt, scale=1}] coordinates {(5, 0.023) (6, 0.024) (7, 0.0206) (8, 0.024) (9, 0.0233)};
    \addplot[mark=triangle*, color=darkorange, mark options={rotate=180,fill=white, draw=darkorange, line width=0.1pt, scale=1}] coordinates {(5, 0.022) (6, 0.0237) (7, 0.021) (8, 0.023) (9, 0.0229)};
    \legend{SCT $L=1$, NPD $L=1$, NPD $L=2$,NPD $L=4$,NPD $L=8$,}
    \end{axis}
\end{tikzpicture}
    \caption{\Glspl{fer} attained by \gls{scl} decoding of the \gls{npd} for $\delta=0.1$. The bar plot shows the information rate per block length.}
    \label{fig:fer-scl-1}
\end{figure}

Figures \ref{fig:fer-scl-01} and \ref{fig:fer-scl-1} shows the performance of a \gls{scl} decoder for $\delta=0.01$ and $\delta=0.1$, respectively. Both figures show the performance of only the \gls{npd}, as the \gls{sct} decoder running time is too high already for the \gls{sc} decoder alone. It is shown, that as expected, incorporating \gls{scl} decoding with the \gls{npd} reduces the obtained \glspl{fer}. 

\section{Conclusion}

\par This work presents a novel approach to decoding deletion channels using \glspl{npd}. Deletion channels, characterized by synchronization errors and modeled by a constant deletion rate, pose significant challenges in reliable decoding due to the loss of synchronization. Existing methods, such as the trellis-based \gls{sc} decoder introduced in \cite{talPolarCodesDeletion2021}, provided a major advancement by enabling the exact computation of posterior probabilities. However, the high computational complexity of $O(N^4)$ limits its practicality for larger block lengths.

\par Motivated by the sparsity observed in the trellis structure, we proposed leveraging \glspl{npd} to compress the trellis representation and reduce computational complexity. By embedding \glspl{nn} within the \gls{sc} framework, \gls{npd} demonstrated two significant advantages: eliminating the need for an explicit channel model and achieving a complexity determined by the \glspl{nn}' parameterization. This led to a reduced computational complexity compared to traditional trellis-based decoders, enabling efficient decoding for longer block lengths and the integration of list decoding.

\par The results demonstrate that the \gls{npd} achieves comparable performance to the trellis-based \gls{sc} decoder while significantly reducing complexity. This advancement not only broadens the applicability of polar codes to deletion channels but also sets the foundation for further exploration into scalable, data-driven decoding techniques for channels with synchronization errors. Future work could focus on extending this approach to other channels with synchronization problems, such as the deletion-insertion-substitution channel. We are also interested on investigating the potential of \glspl{npd} for the DNA storage problem.


\newpage
\balance
\printbibliography

@article{aharoniDatadrivenNeuralPolar2024,
  title = {Data-Driven {{Neural Polar Decoders}} for {{Unknown Channels}} with and without {{Memory}}},
  author = {Aharoni, Ziv and Huleihel, Bashar and Pfister, Henry D and Permuter, Haim H},
  year = {2024},
  journal = {IEEE Transactions on Information Theory},
  publisher = {IEEE}
}

@article{arikanChannelPolarizationMethod2009,
  title = {Channel {{Polarization}}: {{A Method}} for {{Constructing Capacity-achieving Codes}} for {{Symmetric Binary-input Memoryless Channels}}},
  author = {Arikan, E.},
  year = {2009},
  journal = {IEEE Trans. Inf. Theory},
  volume = {55},
  number = {7},
  pages = {3051--3073},
  publisher = {IEEE}
}

@article{bancroftLongtermStorageInformation2001,
  title = {Long-Term Storage of Information in {{DNA}}},
  author = {Bancroft, Carter and Bowler, Timothy and Bloom, Brian and Clelland, Catherine Taylor},
  year = {2001},
  journal = {Science},
  volume = {293},
  number = {5536},
  pages = {1763--1765},
  publisher = {American Association for the Advancement of Science}
}

@article{churchNextgenerationDigitalInformation2012,
  title = {Next-Generation Digital Information Storage in {{DNA}}},
  author = {Church, George M and Gao, Yuan and Kosuri, Sriram},
  year = {2012},
  journal = {Science},
  volume = {337},
  number = {6102},
  pages = {1628--1628},
  publisher = {American Association for the Advancement of Science}
}

@article{daveyReliableCommunicationChannels2001,
  title = {Reliable Communication over Channels with Insertions, Deletions, and Substitutions},
  author = {Davey, Matthew C and MacKay, David JC},
  year = {2001},
  journal = {IEEE Transactions on Information Theory},
  volume = {47},
  number = {2},
  pages = {687--698},
  publisher = {IEEE}
}

@article{dobrushinShannonsTheoremsChannels1967,
  title = {Shannon's Theorems for Channels with Synchronization Errors},
  author = {Dobrushin, Roland L'vovich},
  year = {1967},
  journal = {Problemy Peredachi Informatsii},
  volume = {3},
  number = {4},
  pages = {18--36},
  publisher = {{Russian Academy of Sciences, Branch of Informatics, Computer Equipment and {\dots}}}
}

@book{gallagerSequentialDecodingBinary1961,
  title = {Sequential Decoding for Binary Channels with Noise and Synchronization Errors},
  author = {Gallager, Robert G},
  year = {1961},
  publisher = {British Library, Reports \& Microfilms}
}

@article{goldmanPracticalHighcapacityLowmaintenance2013,
  title = {Towards Practical, High-Capacity, Low-Maintenance Information Storage in Synthesized {{DNA}}},
  author = {Goldman, Nick and Bertone, Paul and Chen, Siyuan and Dessimoz, Christophe and LeProust, Emily M and Sipos, Botond and Birney, Ewan},
  year = {2013},
  journal = {nature},
  volume = {494},
  number = {7435},
  pages = {77--80},
  publisher = {Nature Publishing Group UK London}
}

@article{sloaneSingledeletioncorrectingCodes2002,
  title = {On Single-Deletion-Correcting Codes},
  author = {Sloane, Neil JA},
  year = {2002},
  journal = {Codes and designs},
  volume = {10},
  pages = {273--291},
  publisher = {Walter de Gruyter, Berlin: Ray-Chaudhuri Festschrift}
}

@article{talListDecodingPolar2015,
  title = {List {{Decoding}} of {{Polar Codes}}},
  author = {Tal, I. and Vardy, A.},
  year = {2015},
  journal = {IEEE Trans. Inf. Theory},
  volume = {61},
  number = {5},
  pages = {2213--2226},
  publisher = {IEEE}
}

@article{talPolarCodesDeletion2021,
  title = {Polar {{Codes}} for the {{Deletion Channel}}: {{Weak}} and {{Strong Polarization}}},
  author = {Tal, I. and Pfister, H. D. and Fazeli, A. and Vardy, A.},
  year = {2021},
  journal = {IEEE Trans. Inf. Theory},
  volume = {68},
  number = {4},
  pages = {2239--2265},
  publisher = {IEEE}
}

@article{thomasPolarCodingBinary2017,
  title = {Polar Coding for the Binary Erasure Channel with Deletions},
  author = {Thomas, Eldho K and Tan, Vincent YF and Vardy, Alexander and Motani, Mehul},
  year = {2017},
  journal = {IEEE Communications Letters},
  volume = {21},
  number = {4},
  pages = {710--713},
  publisher = {IEEE}
}

@inproceedings{tianPolarCodesChannels2017,
  title = {Polar Codes for Channels with Deletions},
  booktitle = {2017 55th {{Annual Allerton Conference}} on {{Communication}}, {{Control}}, and {{Computing}} ({{Allerton}})},
  author = {Tian, Kuangda and Fazeli, Arman and Vardy, Alexander and Liu, Rongke},
  year = {2017},
  pages = {572--579},
  publisher = {IEEE}
}

@article{varshamovCodesWhichCorrect1965a,
  title = {Codes Which Correct Single Asymmetric Errors},
  author = {Varshamov, R. R. and Tenengolts, G. M.},
  year = {1965},
  journal = {Automation and Remote Control},
  volume = {26},
  number = {2},
  pages = {286--290}
}

@inproceedings{vaswaniAttentionAllYou2017,
  title = {Attention Is {{All}} You {{Need}}},
  booktitle = {Advances in {{Neural Information Processing Systems}}},
  author = {Vaswani, Ashish and Shazeer, Noam and Parmar, Niki and Uszkoreit, Jakob and Jones, Llion and Gomez, Aidan N and ukasz Kaiser, {\L} and Polosukhin, Illia},
  editor = {Guyon, I. and Luxburg, U. Von and Bengio, S. and Wallach, H. and Fergus, R. and Vishwanathan, S. and Garnett, R.},
  year = {2017},
  volume = {30},
  publisher = {Curran Associates, Inc.}
}

@inproceedings{wangConstructionPolarCodes2015,
  title = {Construction of {{Polar Codes}} for {{Channels}} with {{Memory}}},
  booktitle = {2015 {{IEEE Information Theory Workshop-Fall}} ({{ITW}})},
  author = {Wang, R. and Honda, J. and Yamamoto, H. and Liu, R. and Hou, Y.},
  year = {2015},
  pages = {187--191},
  publisher = {IEEE}
}

@article{zigangirovSequentialDecodingBinary1969,
  title = {Sequential Decoding for a Binary Channel with Drop-Outs and Insertions},
  author = {Zigangirov, Kamil'Shamil'evich},
  year = {1969},
  journal = {Problemy Peredachi Informatsii},
  volume = {5},
  number = {2},
  pages = {23--30},
  publisher = {{Russian Academy of Sciences, Branch of Informatics, Computer Equipment and {\dots}}}
}
\end{document}